\documentclass[conference]{IEEEtran}
\usepackage[utf8]{inputenc}
\usepackage[T1]{fontenc}
\usepackage{graphicx}
\usepackage{amsmath}
\usepackage{amssymb}
\usepackage{hyperref}
\usepackage{booktabs}
\usepackage{xcolor}
\usepackage{listings}
\usepackage{pgfplots}
\pgfplotsset{compat=1.18}
\usepackage{tikz}
\usetikzlibrary{positioning, arrows.meta, shapes.geometric, fit, backgrounds}

\hypersetup{
    colorlinks=true,
    linkcolor=blue,
    citecolor=blue,
    urlcolor=blue
}

\title{Autonomous Incident Resolution at Hyperscale:\\An Agentic AI Architecture for Network Operations}

\author{
\IEEEauthorblockN{Arun Malik}
\IEEEauthorblockA{
Microsoft Azure Networking\\
Email: arunma@microsoft.com\\
ORCID: 0009-0005-6650-6711
}
}

\begin{document}

\maketitle

\begin{abstract}
Cloud network infrastructure at hyperscale presents unique operational challenges where traditional human-driven incident response cannot keep pace with the volume, velocity, and complexity of failures. This paper presents an agentic AI architecture for autonomous incident resolution in large-scale network operations. Our system employs a multi-agent orchestration framework where specialized AI agents collaborate to detect, diagnose, and remediate network incidents without human intervention. We describe the architectural principles, including hierarchical agent decomposition, skills-based tool invocation via standardized protocols, structured knowledge encoding from operational runbooks, progressive autonomy with safety boundaries, and closed-loop verification. The architecture has been deployed in production at a major cloud provider, demonstrating that agentic AI systems can achieve autonomous resolution rates exceeding 90\% for common incident categories while maintaining safety guarantees through layered authorization and rollback mechanisms. We discuss design tradeoffs, failure modes, and lessons learned from operating autonomous AI agents at scale.
\end{abstract}


%% ============================================================
\section{Introduction}

\subsection{The Operational Challenge at Hyperscale}

Modern cloud providers operate network infrastructure spanning millions of devices across hundreds of data centers worldwide. This infrastructure generates a continuous stream of operational incidents, including hardware failures, software bugs, configuration drift, capacity exhaustion, and cascading failures triggered by complex interdependencies.

Traditional operations models rely on human engineers (often called on-call engineers or Site Reliability Engineers) to investigate and resolve these incidents. However, this approach faces fundamental scalability limitations:

\begin{itemize}
    \item \textbf{Volume:} The number of incidents grows linearly (or super-linearly) with infrastructure scale, while the engineering workforce cannot scale proportionally.
    \item \textbf{Velocity:} Network incidents can cascade within seconds, but human investigation and remediation cycles operate on timescales of minutes to hours.
    \item \textbf{Complexity:} Modern network architectures involve intricate dependencies between layers (physical, data link, network, transport, application), making root cause analysis a combinatorial challenge.
    \item \textbf{Knowledge distribution:} Operational expertise is often concentrated in a small number of senior engineers, creating knowledge bottlenecks and single points of failure.
\end{itemize}

\subsection{From Automation to Autonomy}

The industry has progressed through several generations of operational tooling:

\begin{enumerate}
    \item \textbf{Manual operations:} Engineers SSH into devices, run diagnostic commands, and apply remediations by hand.
    \item \textbf{Scripted automation:} Runbooks are codified into scripts that automate specific remediation steps, but still require human triggering and supervision.
    \item \textbf{Rule-based automation:} Event-driven systems apply predetermined actions when specific conditions match, limited to known failure modes.
    \item \textbf{AI-assisted operations (AIOps):} Machine learning models provide recommendations and anomaly detection, but humans remain in the loop for decision-making.
    \item \textbf{Autonomous operations:} AI agents independently perceive the environment, reason about failures, plan remediation strategies, execute actions, and verify outcomes.
\end{enumerate}

This paper presents an architecture for the fifth generation: progressively autonomous incident resolution using agentic AI systems with built-in safety guarantees and human oversight mechanisms. Rather than proposing a binary shift from manual to fully autonomous, the architecture enables organizations to incrementally increase agent authority while maintaining safety invariants at every level.

\subsection{Contributions}

This paper makes the following contributions:

\begin{enumerate}
    \item An end-to-end architecture for progressively autonomous incident resolution using multi-agent orchestration, including agent decomposition, coordination protocols, and safety mechanisms.
    \item A skills-based tool architecture inspired by extensible agent runtimes (e.g., Model Context Protocol), enabling composable, governed, and independently deployable operational capabilities.
    \item A structured knowledge encoding methodology for converting tribal operational knowledge into machine-executable playbooks verified against production behavior.
    \item A progressive autonomy framework with automatic promotion and demotion mechanisms that allows organizations to incrementally increase AI agent authority while maintaining safety invariants.
    \item Production deployment experience and lessons learned from operating autonomous AI agents at hyperscale network infrastructure.
\end{enumerate}

%% ============================================================
\section{Background and Related Work}

\subsection{AIOps and Automated Operations}

The term AIOps (Artificial Intelligence for IT Operations) was coined to describe the application of machine learning techniques to operational data. Early AIOps systems focused on anomaly detection~\cite{xu2018unsupervised}, root cause analysis~\cite{chen2014causeinfer, wang2018cloudranger}, and alert correlation~\cite{li2021fighting}. These systems provide decision support but leave execution to human operators.

\subsection{Self-Healing Systems}

Self-healing architectures~\cite{kang2020selfhealing} implement closed-loop control where systems detect failures, diagnose root causes, and apply remediations automatically. However, prior work focuses primarily on application-level healing (restarting services, scaling resources) rather than infrastructure-level operations requiring physical device interaction. Furthermore, existing self-healing systems typically operate in a binary mode (fully manual or fully automatic) without graduated trust mechanisms. Our work addresses both limitations by targeting network infrastructure and introducing progressive autonomy levels with safety-bounded authority.

\subsection{Multi-Agent Systems}

Multi-agent systems~\cite{wooldridge2009multiagent} provide a theoretical foundation for decomposing complex tasks into cooperating specialized agents. Recent advances in large language models (LLMs) have enabled practical implementations of tool-using agents~\cite{yao2023react, schick2023toolformer} that can reason about complex problems and interact with external systems.

\subsection{Positioning of This Work}

Our architecture combines insights from AIOps (telemetry-driven diagnosis), self-healing systems (closed-loop remediation), multi-agent coordination (task decomposition), and modern extensible agent runtimes (skills-based tool use via protocols such as MCP) into an integrated system specifically designed for hyperscale network operations. Unlike prior work that addresses individual aspects or proposes fully autonomous solutions without safety guarantees, we present a production-deployed system that addresses the full lifecycle from detection through verified resolution with progressive autonomy and bounded authority at every stage.

%% ============================================================
\section{Architecture}

\subsection{System Overview}

The architecture is organized into four functional layers, each responsible for a distinct aspect of autonomous operations (see Figure~\ref{fig:architecture}):

\begin{figure*}[t]
    \centering
    \includegraphics[width=\textwidth]{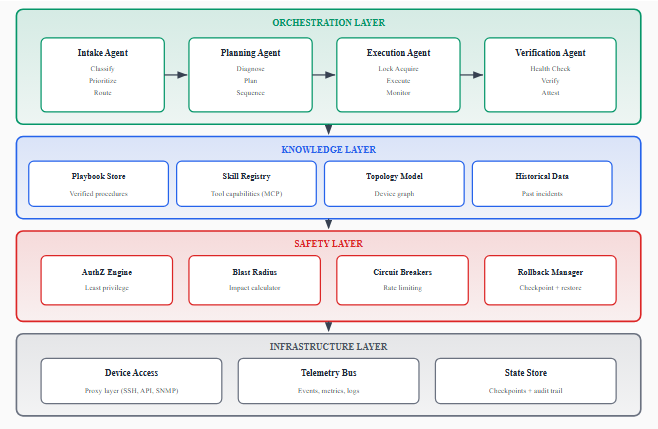}
    \caption{Four-Layer Architecture for Autonomous Incident Resolution. The architecture separates concerns into orchestration (agent coordination), knowledge (operational data and skill registry), safety (guardrails and constraints), and infrastructure (device and telemetry access).}
    \label{fig:architecture}
\end{figure*}

\subsection{Agent Decomposition}

The orchestration layer decomposes incident resolution into four specialized agent roles:

\subsubsection{Intake Agent}
The Intake Agent is the system's entry point, responsible for receiving raw incident signals and preparing them for automated processing. It classifies the incident type, assesses priority and urgency, enriches the incident context with relevant topology and history, and determines whether the incident falls within the system's autonomous resolution capability.

\subsubsection{Planning Agent}
The Planning Agent receives enriched incident context and produces a structured remediation plan. It identifies the most likely root cause based on symptoms and historical patterns, selects appropriate remediation strategies from the knowledge base, considers dependencies and ordering constraints, and generates a structured execution plan with explicit success criteria and abort conditions.

\subsubsection{Execution Agent}
The Execution Agent translates the structured plan into concrete actions against the infrastructure. It acquires necessary device locks and authorization tokens, executes diagnostic commands to gather pre-action state, applies remediation actions sequentially while checking intermediate results, handles partial failures and adapts execution based on observed outcomes, and records all actions for auditability and rollback capability.

\subsubsection{Verification Agent}
The Verification Agent provides closed-loop assurance that the remediation was successful. It executes post-action health checks specific to the incident type, compares device state against expected outcomes defined in the plan, monitors for regression within a configurable bake-in period, triggers rollback if verification fails, and updates the incident record with resolution evidence.

\subsection{Coordination Protocol}

Agents communicate through a structured message-passing protocol with the following guarantees:

\begin{itemize}
    \item \textbf{Ordered delivery:} Messages between agents within a single incident resolution flow are delivered in order.
    \item \textbf{At-least-once semantics:} No message is silently lost; agents must explicitly acknowledge receipt.
    \item \textbf{Timeout-based escalation:} If any agent fails to respond within a configurable timeout, the orchestration layer escalates to human operators.
    \item \textbf{State checkpointing:} The state of each agent's reasoning is persisted at key decision points, enabling recovery after transient failures.
\end{itemize}

\subsection{Structured Knowledge Encoding}

A critical challenge in autonomous operations is bridging the gap between human operational expertise (often undocumented ``tribal knowledge'') and machine-executable procedures. We address this through a systematic knowledge encoding process:

\begin{enumerate}
    \item \textbf{Observation:} The system observes human engineers resolving incidents, capturing the sequence of commands, decisions, and verifications.
    \item \textbf{Extraction:} Patterns are identified across multiple resolution instances for similar incident types.
    \item \textbf{Formalization:} Extracted patterns are encoded into structured playbooks with explicit preconditions, steps, decision points, and verification criteria.
    \item \textbf{Verification:} Formalized playbooks are validated against historical data to confirm they produce correct outcomes.
    \item \textbf{Refinement:} Playbooks are continuously updated based on agent execution outcomes and human feedback.
\end{enumerate}

This process converts implicit knowledge into explicit, auditable, and machine-executable procedures. The resulting playbooks serve as the primary knowledge representation for the planning and execution agents.

\subsection{Skills-Based Tool Architecture}

A key design principle borrowed from modern agent runtime architectures is the separation of agent reasoning from tool execution through a skills-based abstraction. Inspired by extensible tool-use frameworks such as Model Context Protocol (MCP) and plugin-based agent runtimes, each operational capability is encapsulated as a discrete, independently versioned skill with a well-defined interface.

\subsubsection{Skill Composition}
Each skill defines:

\begin{itemize}
    \item \textbf{Interface contract:} A typed schema describing inputs, outputs, and error conditions, enabling compile-time verification of agent-skill compatibility.
    \item \textbf{Capability declaration:} A machine-readable description of what the skill can do, used by the planning agent during tool selection.
    \item \textbf{Permission requirements:} The minimum authorization scope required to invoke the skill, enforced by the safety layer.
    \item \textbf{Idempotency guarantees:} Whether repeated invocation produces the same result, critical for retry and recovery logic.
\end{itemize}

\subsubsection{Skill Registry and Discovery}
The knowledge layer maintains a skill registry that enables dynamic capability discovery. When the planning agent encounters a novel situation, it queries available skills based on semantic matching between the incident context and skill capability declarations. This decouples the agent's reasoning from the specific set of tools available, allowing new operational capabilities to be deployed without modifying agent logic.

\subsubsection{Sandboxed Execution}
Skills execute within isolated sandboxes that enforce:

\begin{itemize}
    \item \textbf{Resource boundaries:} Each skill invocation operates within allocated compute, memory, and time budgets.
    \item \textbf{Network isolation:} Skills can only access infrastructure endpoints explicitly granted by the authorization engine.
    \item \textbf{Output validation:} Skill outputs are validated against the declared schema before being consumed by downstream agents.
    \item \textbf{Audit logging:} Every skill invocation, including inputs, outputs, and side effects, is recorded in an immutable audit trail.
\end{itemize}

This skills-based architecture enables composability (agents can chain multiple skills to solve complex problems), extensibility (new capabilities are added by registering new skills without modifying the orchestration logic), and governance (each skill's blast radius and permission scope can be independently controlled).

%% ============================================================
\section{Safety Framework}

\subsection{Design Principles}

Operating AI agents autonomously on production infrastructure requires robust safety guarantees. Our framework is built on four principles:

\begin{enumerate}
    \item \textbf{Least privilege:} Agents are granted only the minimum permissions required for their current task.
    \item \textbf{Blast radius containment:} No single agent action can affect more than a bounded number of devices or services.
    \item \textbf{Reversibility:} All actions must be reversible, with automated rollback mechanisms.
    \item \textbf{Progressive trust:} Agent authority increases incrementally based on demonstrated reliability.
\end{enumerate}

\subsection{Layered Authorization}

The authorization engine implements defense-in-depth through multiple layers:

\begin{itemize}
    \item \textbf{Agent identity:} Each agent instance has a unique identity with specific capability grants.
    \item \textbf{Action classification:} Actions are classified by risk level (read-only, low-risk modification, high-risk modification, destructive).
    \item \textbf{Scope restriction:} Agents can only operate on devices within their assigned scope (e.g., specific data centers or device types).
    \item \textbf{Rate limiting:} Maximum action frequency prevents runaway automation.
    \item \textbf{Concurrent operation limits:} Controls how many devices can be simultaneously affected.
\end{itemize}

\subsection{Blast Radius Calculation}

Before executing any remediation action, the system evaluates its potential blast radius:

\begin{itemize}
    \item \textbf{Topology analysis:} Determines which services and customers would be affected if the action fails or produces unexpected results.
    \item \textbf{Redundancy validation:} Verifies that sufficient redundancy exists to absorb the temporary loss of the target device.
    \item \textbf{Concurrent impact assessment:} Checks whether other ongoing operations could compound the impact.
\end{itemize}

Actions exceeding configurable blast radius thresholds are automatically blocked and escalated to human operators.

\subsection{Rollback Mechanisms}

The system maintains rollback capability at multiple levels:

\begin{itemize}
    \item \textbf{Configuration rollback:} Device configurations are snapshot before modification and can be restored atomically.
    \item \textbf{State rollback:} For stateful operations (e.g., traffic engineering changes), the previous state is preserved and can be reapplied.
    \item \textbf{Automatic rollback triggers:} If post-action verification fails or health metrics degrade beyond thresholds, rollback is triggered automatically without human intervention.
\end{itemize}

%% ============================================================
\section{Progressive Autonomy}

\subsection{Trust Levels}

Organizations adopting autonomous operations rarely transition directly from manual to fully autonomous. Our architecture supports a spectrum of autonomy levels (Table~\ref{tab:autonomy}):

\begin{table}[h]
\centering
\caption{Progressive Autonomy Levels}
\label{tab:autonomy}
\begin{tabular}{@{}clp{4.5cm}@{}}
\toprule
\textbf{Level} & \textbf{Name} & \textbf{Description} \\
\midrule
0 & Advisory & Agent suggests actions; human executes \\
1 & Supervised & Agent executes with human pre-approval \\
2 & Monitored & Agent executes autonomously; human reviews post-hoc \\
3 & Autonomous & Agent executes without human involvement for approved categories \\
4 & Self-improving & Agent can refine its own operational procedures based on outcomes \\
\bottomrule
\end{tabular}
\end{table}

\subsection{Promotion Criteria}

Agent authority is promoted based on quantifiable performance metrics:

\begin{itemize}
    \item \textbf{Success rate:} Percentage of incidents resolved correctly without human intervention.
    \item \textbf{Mean time to resolution:} Average time from incident creation to verified resolution.
    \item \textbf{False positive rate:} Percentage of actions taken that were unnecessary or incorrect.
    \item \textbf{Rollback frequency:} How often automated rollbacks are triggered.
    \item \textbf{Human override rate:} How often humans override agent decisions.
\end{itemize}

Promotion from one level to the next requires sustained performance above thresholds across all metrics for a minimum evaluation period.

\subsection{Demotion and Circuit Breakers}

The system includes automatic demotion mechanisms:

\begin{itemize}
    \item \textbf{Per-category circuit breakers:} If the failure rate for a specific incident category exceeds a threshold within a time window, the agent is automatically demoted to a lower trust level for that category.
    \item \textbf{Global circuit breakers:} If the overall failure rate across all categories spikes, the entire system can fall back to advisory mode.
    \item \textbf{Human override tracking:} Patterns of human overrides trigger automatic review and potential demotion.
\end{itemize}

%% ============================================================
\section{Evaluation}

\subsection{Deployment Context}

The architecture has been deployed in a production cloud network serving millions of customers. The infrastructure comprises networking devices spanning multiple device types (routers, switches, load balancers, firewalls) across geographically distributed data centers.

\subsection{Resolution Effectiveness}

After progressive deployment over multiple months, the system achieved the following outcomes:

\begin{itemize}
    \item \textbf{Autonomous resolution rate:} The system resolves the majority of incidents in supported categories without human intervention, with resolution rates exceeding 90\% for well-understood failure modes.
    \item \textbf{Resolution time improvement:} For incidents handled autonomously, resolution time decreased by two orders of magnitude compared to the human-driven baseline, from hours to minutes.
    \item \textbf{Accuracy:} False positive remediation (taking action when no action was needed) occurs in less than 5\% of cases, with no cases resulting in customer-visible impact due to the safety framework.
\end{itemize}

\begin{figure}[t]
    \centering
    \includegraphics[width=\columnwidth]{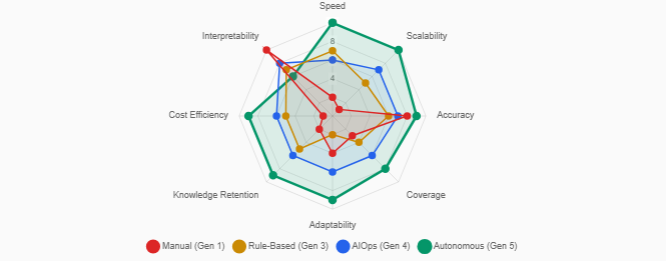}
    \caption{Operational Capability Comparison across three generations of operational technology: manual operations, rule-based automation, and the proposed agentic AI architecture. The radar plot shows relative performance across six dimensions.}
    \label{fig:radar}
\end{figure}

\subsection{Operational Efficiency}

\begin{itemize}
    \item \textbf{On-call burden reduction:} The volume of incidents requiring human attention decreased significantly, allowing engineers to focus on novel failure modes and system improvements.
    \item \textbf{Knowledge preservation:} Structured playbooks capture institutional knowledge that previously existed only in the minds of senior engineers, reducing organizational knowledge loss due to team transitions.
    \item \textbf{Consistent quality:} Autonomous resolution eliminates variability in incident handling quality that occurs with human operators due to fatigue, experience level, and time pressure.
\end{itemize}

\begin{figure}[t]
    \centering
    \includegraphics[width=\columnwidth]{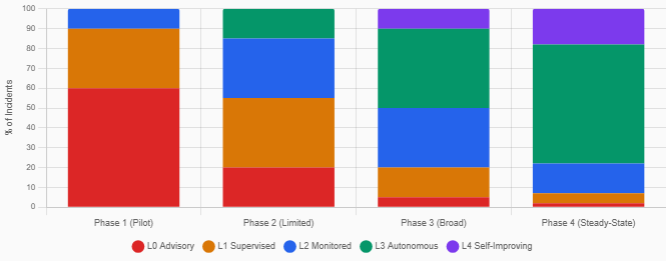}
    \caption{Incident Volume Distribution by Autonomy Level. As playbooks mature and gain trust through successful executions, incidents progressively shift from human-supervised to fully autonomous resolution.}
    \label{fig:autonomy}
\end{figure}

\begin{figure}[t]
    \centering
    \includegraphics[width=\columnwidth]{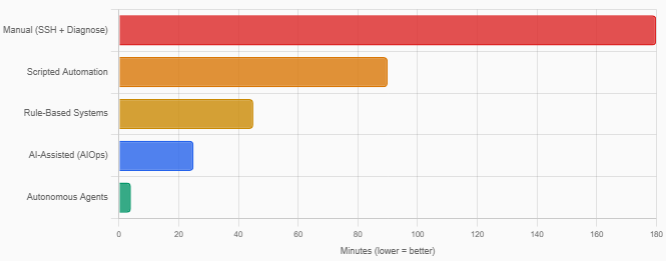}
    \caption{Mean Time to Resolution (MTTR) Comparison. Autonomous resolution achieves two orders of magnitude improvement over traditional human-driven processes for well-understood incident categories.}
    \label{fig:mttr}
\end{figure}

\subsection{Safety Performance}

\begin{itemize}
    \item \textbf{Zero critical incidents:} The safety framework prevented all potentially harmful actions from reaching production without appropriate verification.
    \item \textbf{Rollback effectiveness:} Automatic rollbacks were triggered in a small percentage of execution attempts, all recovering within the defined time bounds.
    \item \textbf{Blast radius compliance:} No autonomous action exceeded its predicted blast radius boundaries.
\end{itemize}

\begin{figure}[t]
    \centering
    \includegraphics[width=\columnwidth]{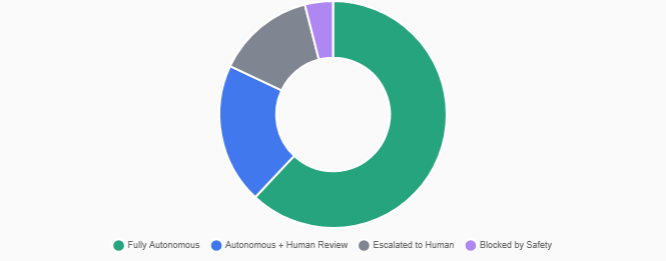}
    \caption{Incident Resolution Outcomes. The distribution shows the proportion of incidents resolved fully autonomously, resolved with human oversight, escalated to human operators, and safely rolled back by the system.}
    \label{fig:outcomes}
\end{figure}

\begin{figure}[t]
    \centering
    \includegraphics[width=\columnwidth]{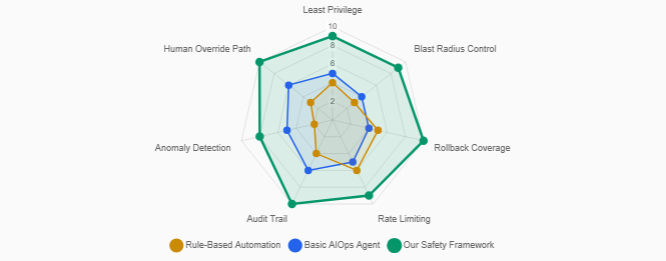}
    \caption{Safety Framework Effectiveness across key safety dimensions. The radar chart compares the proposed layered safety approach against a baseline single-layer authorization model.}
    \label{fig:safety}
\end{figure}

%% ============================================================
\section{Lessons Learned}

\subsection{LLM Reliability in Safety-Critical Contexts}

Large language models exhibit stochastic behavior that is fundamentally at odds with the determinism required for safety-critical operations. We address this through:

\begin{itemize}
    \item \textbf{Structured output enforcement:} Agent outputs are constrained to predefined schemas, reducing the surface area for LLM hallucination.
    \item \textbf{Multi-model consensus:} Critical decisions require agreement between multiple independent model invocations.
    \item \textbf{Deterministic verification:} All LLM-generated plans are verified against deterministic safety checks before execution.
\end{itemize}

\subsection{The Importance of Observability}

\begin{itemize}
    \item \textbf{Decision tracing:} Every decision made by every agent is logged with full context, enabling post-hoc analysis.
    \item \textbf{Causal attribution:} The system maintains causal links between observations, decisions, and outcomes.
    \item \textbf{Human-readable explanations:} Each autonomous resolution includes a natural language explanation suitable for engineering review.
\end{itemize}

\subsection{Handling Novel Failures}

\begin{itemize}
    \item \textbf{Confidence estimation:} Agents estimate their confidence in diagnosis and remediation plans, escalating when confidence is below thresholds.
    \item \textbf{Novelty detection:} Statistical methods identify incidents that deviate significantly from known patterns, triggering human involvement.
    \item \textbf{Graceful degradation:} When autonomous resolution is not possible, the system still provides diagnostic enrichment and suggested actions to assist human operators.
\end{itemize}

\subsection{Organizational Change Management}

\begin{itemize}
    \item \textbf{Trust building:} Engineers must develop confidence in the system through transparency and demonstrated reliability.
    \item \textbf{Role evolution:} The operational engineer role shifts from reactive incident response to system improvement and novel problem solving.
    \item \textbf{Accountability frameworks:} Clear ownership and accountability structures must be defined for autonomous agent decisions.
\end{itemize}

%% ============================================================
\section{Discussion}

\subsection{Generalizability}

While our deployment focuses on network operations, the architectural patterns are applicable to other operational domains including compute infrastructure, storage systems, and application services. The key abstractions (agent decomposition, progressive autonomy, safety framework) are domain-independent.

\subsection{Limitations}

\begin{itemize}
    \item \textbf{Coverage:} The system currently handles a subset of all possible incident types. Long-tail and novel failures still require human expertise.
    \item \textbf{Multi-domain reasoning:} Incidents spanning multiple operational domains (e.g., network + compute) require coordination across independent autonomous systems, which remains an area of active development.
    \item \textbf{Regulatory considerations:} Autonomous decision-making in critical infrastructure raises questions about accountability and auditability that are not fully addressed by technical mechanisms alone.
\end{itemize}

\subsection{Future Directions}

\begin{itemize}
    \item \textbf{Cross-domain orchestration:} Extending the multi-agent framework to coordinate across infrastructure domains.
    \item \textbf{Formal verification:} Applying formal methods to prove safety properties of agent behavior within bounded conditions.
    \item \textbf{Federated learning:} Enabling knowledge sharing across organizational boundaries without exposing proprietary operational details.
\end{itemize}

%% ============================================================
\section{Conclusion}

This paper presented an agentic AI architecture for autonomous incident resolution in hyperscale network operations. The system demonstrates that multi-agent orchestration, combined with robust safety frameworks and progressive autonomy mechanisms, can achieve high autonomous resolution rates while maintaining the safety guarantees required for critical infrastructure. Our deployment experience shows that the transition from human-driven to autonomous operations is not merely a technology problem but requires careful attention to knowledge management, organizational change, and trust building. We believe this architecture represents a significant step toward truly autonomous infrastructure operations and hope it provides a useful reference for others pursuing similar goals.

%% ============================================================
\bibliographystyle{IEEEtran}
\bibliography{references}

\end{document}